\documentclass[aps,prb,twocolumn,showpacs,superscriptaddress,10pt,longbibliography]{revtex4-2}

\usepackage[colorlinks = true,linkcolor = red,citecolor = magenta]{hyperref}
\usepackage[sort&compress]{natbib}
\usepackage{scalerel}
\usepackage[normalem]{ulem}
\usepackage{xcolor}
\usepackage{bm}
\usepackage{amsmath}
\usepackage{amsfonts}
\usepackage{amssymb}
\usepackage{graphicx}
\usepackage{subfigure}
\usepackage{mathtools}
\usepackage{physics}
\usepackage{dsfont}
\usepackage{float}
\usepackage{lmodern}
\usepackage{empheq}
\usepackage{cleveref}
\usepackage{textcomp}
\usepackage{soul}

\newcommand{\wl}{\mathcal{W}}
\newcommand{\msp}{\mathcal{S}}
\newcommand{\mh}{\mathcal{H}}
\newcommand{\mz}{\mathbb{Z}}
\newcommand{\mr}{\mathbb{R}}
\newcommand{\mF}{\mathcal{F}}
\newcommand{\lgn}{\Delta}

\newcommand{\pcsadd}{Center for Theoretical Physics of Complex Systems, Institute for Basic Science (IBS), Daejeon 34126, Republic of Korea}
\newcommand{\ustadd}{Basic Science Program, Korea University of Science and Technology (UST), Daejeon 34113, Korea}

\begin{document}

\title{Intermediate super-exponential localization with Aubry-Andr\'e chains}

\author{Arindam Mallick}
   \email{arindam.mallick@uj.edu.pl}
   \affiliation{\pcsadd}
   \affiliation{Instytut Fizyki Teoretycznej, Uniwersytet Jagiello\'nski, Lojasiewicza 11, 30-348 Krak\'ow, Poland}

\author{Alexei Andreanov}
   \email{aalexei@ibs.re.kr}
   \affiliation{\pcsadd}
   \affiliation{\ustadd}

\author{Sergej Flach}
   \email{sflach@ibs.re.kr}
   \affiliation{\pcsadd}
   \affiliation{\ustadd}

\date{\today}

\begin{abstract}
  We demonstrate the existence of an intermediate super-exponential localization regime for eigenstates of the Aubry-Andr\'e chain.
  In this regime, the eigenstates localize factorially similarly to the eigenstates of the Wannier-Stark ladder.
  The super-exponential decay emerges on intermediate length scales for large values of the \emph{winding length}---the quasi-period of the Aubry-Andr\'e potential.
  This intermediate localization is present both in the metallic and insulating phases of the system. 
  In the insulating phase, the super-exponential localization is periodically interrupted by weaker decaying tails to form the conventional asymptotic exponential decay predicted for the Aubry-Andr\'e model.
  In the metallic phase, the super-exponential localization happens for states with energies away from the center of the spectrum and is followed by a super-exponential growth into the next peak of the extended eigenstate.
  By adjusting the parameters it is possible to arbitrarily extend the validity of the super-exponential localization.
  A similar intermediate super-exponential localization regime is demonstrated in quasiperiodic discrete-time unitary maps.
\end{abstract}

\maketitle

\section{Introduction}

Uncorrelated disorder potentials enforce exponential Anderson localization for all eigenstates in one-dimensional potentials~\cite{anderson1958absence,kramer1993localization}. 
This holds for any strength of the disorder potential relative to the kinetic energy strength induced e.g.~by the hopping \(t\) between nearest neighbour (n.n.) sites on a one-dimensional chain.
Quasi-periodic potentials instead can be viewed as highly correlated aperiodic ones.
Consequently the one-dimensional model shows a metal-insulator transition with eigenstates changing from extended to exponentially localized upon increasing the quasi-periodic potential strength \(\lambda\) relative to the hopping \(t\).
This was first demonstrated for the Aubry-Andr\'e chain (AA)~\cite{aubry1980analyticity, lahini2009observation}. 
The AA model and its generalizations proved to be useful in different branches of physics, e.g., many-body localization~\cite{mastropietro2015localization, doggen2019many, strkalj2021many, sarkar2022signatures}, and topological insulators~\cite{zeng2020topological, zeng2020higher, fraxanet2021topological}.
Theory predicts~\cite{aubry1980analyticity} and simulations confirm~\cite{lahini2009observation} the asymptotic exponential decay of the eigenstates~\cite{roati2008anderson} for strong enough potential \(\lambda\). 
However the intermediate but potentially long-lasting non-asymptotic decay, i.e.~the decay of the probability profile on intermediate length scales can show substantial deviation from the established asymptotic ones.

The AA model has four parameters: hopping strength \(t\), potential strength \(\lambda\), an irrational \emph{winding length} \(\wl\) or spatial quasi-period of the potential, and a phase \(\phi\) which is usually absorbed by a shift of space origin. 
The AA model is known to show asymptotic exponential eigenstate localization in its localized phase \(\abs{\lambda} > 2\abs{t}\). 
The localization length \(\xi\) is depending only on one parameter---the ratio of potential strength to hopping strength (see the section below). 
In particular, \(\xi\) is not depending on neither the winding length, nor on the position of the eigenstate, nor on its eigenenergy. 
However, the precise form of the localized eigenstate does depend on the winding length, and also on its position in space relative to the AA potential, and also on its energy. 
We show that such eigenstates develop a broad core of nondecaying oscillating parts for large winding length, due to the slow variation of the AA potential and the finite strength of the hopping \(t\). 
At the same time the decay away from the core is controlled over large length scales of the order of the winding length by faster than exponential---super-exponential---factorial decay in the same regime. 
Further details can also depend on the energy and the spatial location of the eigenstate. 
Exponential decay is observed on distances larger than the winding length. 
Remarkably, these features are also observed in the metallic phase, \(\abs{\lambda} < 2\abs{t}\), for states with energies close to the edge of the spectrum. 
The super-exponential decay is similar to the decay of the eigenstates of the Wannier-Stark ladder~\cite{fukuyama1973tightly, mallick2021wannier}.
Very high precision and very large lattice sizes are required in this regime in order to see that such eigenstates are actually exponentially localized or extended, as predicted by the theory~\cite{aubry1980analyticity}.

We further show that similar results hold for unitary maps as well. 
Time evolution under discrete-time unitary maps is usually computationally faster compared to the Hamiltonian evolution~\cite{vakulchyk2018almost, vakulchyk2019wave, malishava2022lyapunov}.
Therefore identifying a unitary map with properties identical or similar to a given Hamiltonian model is important for the development of fast simulation tools for unitary dynamics.
We consider a unitary map (UM): a split-step discrete-time quantum walk~\cite{hamza2009dynamical, kitagawa2012topological, mallick2016dirac} with a spatially varying quasi-periodic phase factor. 
Similarly to the AA Hamiltonian model with \(\abs{\lambda} > 2 \abs{t}\), the eigenstates of this UM localize exponentially~\cite{cedzich2021anderson}.
For smaller phase factor and coin parameters, we identify the intermediate super-exponential localization regime. 
This allows in principle to use these maps to approximate Wannier-Stark Hamiltonians in the unitary map setting and achieve faster simulation of their dynamics.

The article is organized as follows. 
In Sec.~\ref{sec:hamil} we discuss the AA Hamiltonian model and its intermediate localization properties for large winding length \(\wl\), including analytical derivation of the asymptotic decay of eigenstates for power-law potentials on a chain.
Sec.~\ref{sec:num-res} covers numerical evidences for the intermediate super-exponential localization in the AA chain with large winding lengths.
In Sec.~\ref{sec:uc} we construct the quasiperiodic discrete-time unitary map, that features the transient plane wave and the super-exponential localization regimes for fine-tuned parameters, similarly to the AA Hamiltonian.
We conclude in Sec.~\ref{sec:concl}.

\section{Intermediate localization regime in the Aubry-Andr\'e chain}
\label{sec:hamil}

The AA Hamiltonian with an incommensurate potential on the 1D chain reads~\cite{aubry1980analyticity}
\begin{align}
  \mh = \sum_{n \in \mz} [-t(\ketbra{n}{n+1} + \ketbra{n}{n-1}) + \lambda \cos(\alpha n + \phi) \ketbra{n}{n} ]\;.
  \label{eq:full_aa_ham}
\end{align}
Here \(t \in \mr\) is the n.n.~hopping parameter, while \(\alpha/2\pi\) is an irrational number ensuring the quasiperiodicity of the potential. 
Without losing generality we consider \(\alpha > 0\).
The potential \(\cos(\alpha n + \phi)\) almost repeats itself at lattice sites \(n\) and \(n + \lfloor2\pi/\alpha\rfloor\) (or \(n + \lceil2\pi/\alpha\rceil\)) for arbitrary \(n\). 
Therefore we define the \emph{winding length} 
\begin{align}
  \wl = \lceil 2\pi/\alpha \rceil \geq 1\;.                                                                                                                                                                                                                            
\end{align}
The Hamiltonian~\eqref{eq:full_aa_ham} has \emph{asymptotically} exponentially localized eigenstates for \(\abs{\lambda} > 2\abs{t}\), which are peaked around some eigenstate dependent lattice site \(n_0\).
The asymptotic localization length
\begin{align}
  \label{eq:ll}
  \xi = \frac{1}{\ln\abs{\frac{\lambda}{2t}}}
\end{align}
describes the asymptotic decay of all eigenstates \(\psi_n \sim {\rm e}^{-|n|/\xi}\) and is independent of the winding length \(\wl\) and the phase \(\phi\)~\cite{aubry1980analyticity}.

\subsection{Large winding length}

We now demonstrate that there is an \emph{intermediate localization} regime whose span is governed by the winding length \(\wl\) and the phase \(\phi\).
In this intermediate regime eigenfunctions decay faster than exponential similarly to the eigenstates of the Wannier-Stark ladder.
For that we consider the case of very small values of \(\abs{\alpha} \ll 1\): 
there is a region of the 1D chain, a subset of the chain sites \(\mathbb{L} \subset \mz\), such that for all \(n \in \mathbb{L}\) and a particular eigenstate peak \(n_0 \in \mathbb{L}\) the following condition holds
\begin{align}
  \abs{\alpha (n - n_0)}\ll 1\;.
\end{align}
Therefore we can approximate the cosine potential up to the second order expansion in \(\alpha (n - n_0)\) as
\begin{align}
  \cos(\alpha n + \phi) & \approx  
  \cos(\alpha n_0 + \phi) \left[1 - \frac{\alpha^2}{2}{(n - n_0)}^2 \right] \notag \\
  & -\sin(\alpha n_0 + \phi) \left[ \alpha(n-n_0)\right]\;. 
  \label{eq:aa_pot_term}
\end{align}
The choice of the phase \(\phi\) allows to approximate either linear or quadratic potentials. 
For \(\alpha n_0 + \phi \approx (2m+1)\pi/2\) with \(m\in \mz\), the Hamiltonian~\eqref{eq:full_aa_ham} is approximated by the Wannier-Stark Hamiltonian (WS) on a 1D chain
\begin{align}
  \mh = -t\sum_{n \in \mz}  \ketbra{n}{n+1} + \ketbra{n}{n-1} \notag \\
  -{(-1)}^m \lambda \alpha \sum_{n\in \mz} (n - n_0) \ketbra{n}{n}\;. 
  \label{eq:sin_poly1}
\end{align}
The effective DC field strength is given by \(\mF=\lambda\alpha\) which has to be compared to the hopping strength \(t\)~\cite{mallick2021wannier}. 
Given \(\abs{\alpha}\ll 1\), one needs strong potential strengths \(\abs{\lambda/t}\gg 1\) to achieve even moderate field values \(\mF \approx t\).

For \(\alpha n_0 + \phi \approx m\pi\) with \(m\in \mz\) the AA Hamiltonian~\eqref{eq:full_aa_ham} approximates the discrete simple harmonic oscillator Hamiltonian (SHO) on a 1D chain
\begin{align}
  \mh & = - t  \sum_{n \in \mz} \ketbra{n}{n+1} + \ketbra{n}{n-1} \notag \\
  & + {(-1)}^m \lambda \sum_{n\in \mz}\left[1 - \frac{\alpha^2}{2}{(n - n_0)}^2 \right] \ketbra{n}{n}\;. 
  \label{eq:cos_poly2}
\end{align}
The harmonic potential strength is given by \(\mF = \lambda\alpha^2/2\).

\subsection{Asymptotic decay of eigenfunctions for power-law potentials on a chain}
\label{sec:pol-pot}

We start by discussing the localization properties of eigenstates in tight-binding chains with power-law potentials.
A tight-binding lattice Hamiltonian with a simple power potential with exponent \(\mu\) takes the form
\begin{align}
  \mh = \sum_{n \in \mz} \left[-t (\ketbra{n}{n+1} + \ketbra{n}{n-1}) + \mF n^\mu \ketbra{n}{n}\right]\;.
  \label{eq:poly_ham}
\end{align}
The eigenvalue equation for an eigenstate at energy \(E\) reads as
\begin{align}
  E \psi_{n} = -t \psi_{n + 1} - t \psi_{n - 1} + \mF n^\mu \psi_n\;.
  \label{eq:poly_eig}
\end{align}
The eigenstates localize factorially, i.e.~super-exponentially in \(n\) for the linear potential \(\mu=1\), e.g.~the Wannier-Stark case~\cite{fukuyama1973tightly, mallick2021wannier}.
This implies that the ratio \(\psi_n/\psi_{n-1}\) decays asymptotically as \(n^{-1}\) for large \(n\), e.g.~far away from the peak of the eigenstate. 
For larger exponents (\(\mu > 1\)), we expect an even stronger decay with \(n\), and we use this observation to approximate the eigenvalue equation for very large \(n\), away from the peak of the localized eigenstate.
Hence we neglect \(E \psi_n\) and \(t\psi_{n + 1}\) compared to the other terms in Eq.~\eqref{eq:poly_eig} for a finite eigenvalue \(E\), a finite hopping strength \(t\) and a finite field strength \(\mF\), 
and simplify the eigenequation to a simple recursion
\begin{align}
  \frac{t}{\mF} n^{-\mu} \psi_{n - 1} \approx \psi_n\;.
\end{align}
This resolves into
\begin{align}
  \label{eq:decay_prob}
  \psi_{n + n_0} \approx {\left(\frac{t}{\mF}\right)}^n {(n!)}^{-\mu} \psi_{n_0}\;,
\end{align}
where \(n_0\) is the peak of the eigenstate.
Similar arguments apply for very large negative values of \(n\), i.e., on other side of the peak of an eigenstate \(\psi_n\).

It follows from the normalization condition of the probability distribution that \(\psi_{n_0}\) is a finite constant. 
Therefore the eigenstate decays super-exponentially for large positive lattice sites \(n\) as
\begin{align}
  {\left(\frac{t}{\mF}\right)}^n {(n!)}^{-\mu} \approx {\left(\frac{t n^{-\mu}}{\mF e^{-\mu}}\right)}^n {(2\pi n)}^{-\frac{\mu}{2}} \approx e^{-\mu n \ln(n)}\;.
\end{align}
Here the Stirling's approximation \(n! \approx \sqrt{2\pi n}{(n/e)}^n \) for large \(n\) was used.
Therefore we expect the same super-exponential decay of the eigenstates for any power-law potential \(n^\mu\), up to the factor \(\mu\) in the exponent.

We note that this asymptotic behavior is very different from the asymptotic localization properties in the continuous 1D space \(\{x \in \mr\}\) where the eigenvalue equation is
\begin{align}
  E \psi_x = - \frac{d^2\psi_x}{dx^2} + \mF(x-x_0)^\mu \psi_x\;.
\end{align}
For the linear case \(\mu = 1\) the wave function \(\psi_x\) takes the form of Airy function which for \(x \gg x_0\) decays as \(\left(\mF^{1/3} x\right)^{-1/4} e^{-\frac{2}{3} \left(\mF^{1/3} x\right)^\frac{3}{2}}\). 
On the other hand for the simple harmonic potential \(\mu = 2\), the wave function \(\psi_x\) is a product of a Gaussian and Hermite polynomials, their asymptotic decay is controlled by the Gaussian: \(e^{-\frac{\sqrt{\mF} x^2}{2}}\).
In both cases the asymptotic decay is faster than that of the discrete cases, \(e^{-\mu x \ln(x)}\).

\section{Numerical results}
\label{sec:num-res}

We refer to \(\abs{\psi_n}^2\) (modulus squared of an eigenstate's amplitude) as probability at a site \(n\) and talk about probability profile \(\{\abs{\psi_n}^2\}\) of an eigenstate.
As well we refer to the core of a localized eigenstate as its peak. 
We set the hopping strength \(t=1\) and measure \(\alpha\) in units of \(2\pi \varphi\) with the golden ratio \(\varphi=(1+\sqrt{5})/2\approx 1.61803\).
We consider finite chains with open boundary condition, numerically diagonalize the Aubry-Andr\'e Hamiltonian~\eqref{eq:full_aa_ham} with \(\phi = \pi/2\) for different winding lengths \(\wl\) and check the decay of probability profiles of the eigenstates.

To quantify the decay of the probability profiles and discriminate between their exponential and super-exponential decay, we use two metrics.
First, we define the local gradient of the logarithms of the probability with respect to the lattice position
\begin{align}
  \lgn(n) = \frac{\ln\abs{\psi_{n+1}}^2 - \ln\abs{\psi_{n-1}}^2}{2} = \ln\abs{\frac{\psi_{n+1}}{\psi_{n-1}}} \;.
  \label{eq:grad_log_prob}
\end{align}
We refer to it as simply the \emph{gradient} in the text below.
A constant in \(n\) value of the gradient indicates exponential localization while the decay with \(n\) signals faster than exponential decay.
For an exponential decay \(e^{-2|n|/\xi}\) with a constant localization length \(\xi\), \(\lgn (n)\) is a step function, 
with values \(\frac{2}{\xi}\) to the left and \(-\frac{2}{\xi}\) to right of the probability peak \(n=0\).
Second, for a super-exponential e.g.~factorial decay of the eigenstates of the Wannier-Stark ladder the following ratio takes an asymptotically constant value:
\begin{align}
  R(\mu, n) = \frac{\ln\abs{\psi_{|n| + n_0}}^2 - \ln\abs{\psi_{n_0}}^2}{\mu \ln(|n|!) - |n| \ln|t/\mF|} \xrightarrow[n\to\infty]{} - 2
  \label{eq:ratio_log_prob}
\end{align}
as follows from Eq.~\eqref{eq:decay_prob}.
We use the gradient \(\lgn\) and the \(R\) value in our numerical simulations to detect and quantify the super-exponential decay.

In what follows we refer to \emph{localization on intermediate scales} as simply \emph{localization}, while the usual AA exponential localization is referred to as \emph{asymptotic localization}.
In all the other cases the type of localization should be clear from the context.

\subsection{Insulating regime: \(\abs{\lambda} > 2\abs{t}\)}

\begin{figure}
  \includegraphics[width = 0.86\columnwidth]{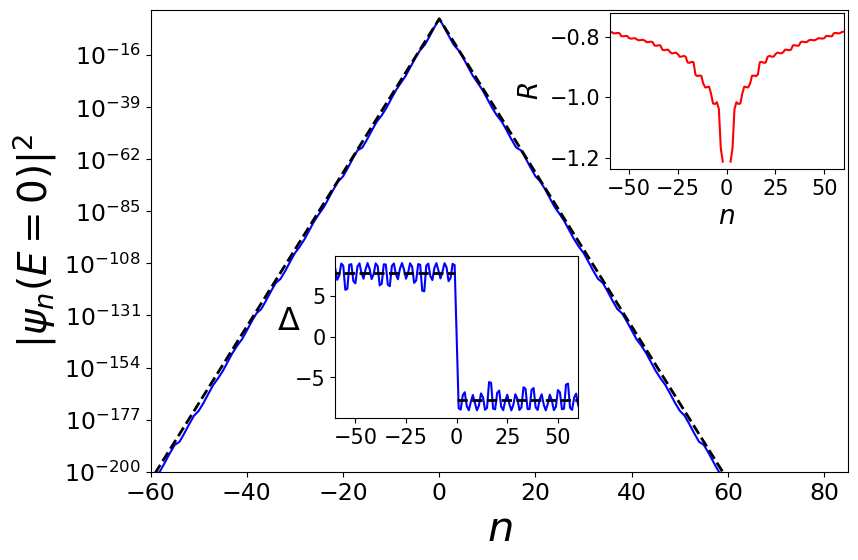}
  \caption{
    The log-linear plot of the spatial profile (blue) of the \(E = 0\) eigenstate of the Hamiltonian~\eqref{eq:full_aa_ham} with \(\wl = 1, (\alpha = 2\pi \varphi\)), \(\lambda = 100\), \(\phi = \pi/2\), 
    \(n_0 = 0\) and \(N=201\) sites with open boundary conditions.
    The black dashed line is the exponential fit with the exponent \(2/\xi = 7.82\) predicted by the theory~\eqref{eq:ll}.
    \emph{Bottom inset}: \(\lgn\)~\eqref{eq:grad_log_prob} (blue) and the same exponential fit (dashed black).
    \emph{Top inset}: \(R\) value~\eqref{eq:ratio_log_prob} evaluated for \(\mF = \lambda \alpha\).
  } 
  \label{fig:sin_ws_exp}
\end{figure}

We remind that our main focus \(\wl \gg 1\). 
Deep in the insulating regime \(\abs{\lambda} \gg 2\abs{t}\) a typical eigenstate with energy close to zero will be located close to a zero of the AA potential and 
have a span (core) 
\begin{gather}
  \label{eq:span}
  \msp \approx \frac{4 \abs{t}}{\abs{\lambda} \alpha} \approx \frac{2\abs{t}}{\pi \abs{\lambda}} \wl, 
\end{gather}
so that \(\msp \ll \wl\) in accord with Eq.~\eqref{eq:sin_poly1}.
There are three lengthscales present in the problem: the localization length \(\xi\), the winding length \(\wl\) and the span of an eigenstate \(\msp\).
The span \(\msp\) is the lengthscale over which the change in potential is of the order of the free particle bandwidth or maximum change in kinetic energy \(4\abs{t}\), 
so that the Hamiltonian appears as that of a free particle on a chain over this lengthscale. 
The eigenstate shows no significant decay features over that length scale. 
Beyond this core the eigenstate will decay super-exponentially until the slow quasiperiodic potential returns to its value at the core center.

In this regime all eigenstates are asymptotically exponentially localized, however their intermediate behavior is controlled and depends strongly on the value of the winding length \(\wl\). 
The parameters of the Hamiltonian~\eqref{eq:full_aa_ham} used below -- \(\lambda = 100, \phi = \pi/2, t = 1\) -- are chosen to approximate the Wannier-Stark Hamiltonian~\eqref{eq:sin_poly1} around lattice position \(n_0 = 0\).
For this choice of parameters the potential energy function~\eqref{eq:full_aa_ham} turns into \(\sin(\alpha n)\).

For small winding length \(\wl=1\), exponential localization is observed on all scales as shown in Fig.~\ref{fig:sin_ws_exp} and the profile of the \(E=0\) eigenstate (blue line) matches well with the exponential decay predicted by the theory (black dashed line).
The estimate of the size of the eigenstate's core~\eqref{eq:span} gives a small value \(\msp \approx 0.004\) which is consistent with the eigenstate being strongly peaked at a single site.
The exponential decay is further confirmed by the behavior of the gradient \(\lgn\) shown in the bottom inset of Fig.~\ref{fig:sin_ws_exp}.
The step-function like behavior of the envelope of the oscillating profile for \(\lgn\) implies exponential decay of the probability profile.
At the same time the \(R\) value~\eqref{eq:ratio_log_prob} evaluated for \(\mF = \lambda \alpha\) is not constant and is different from the \(-2\) value indicating the absence of the super-exponential localization. 

\begin{figure}
  \subfigure[]{\includegraphics[width = 0.9\columnwidth]{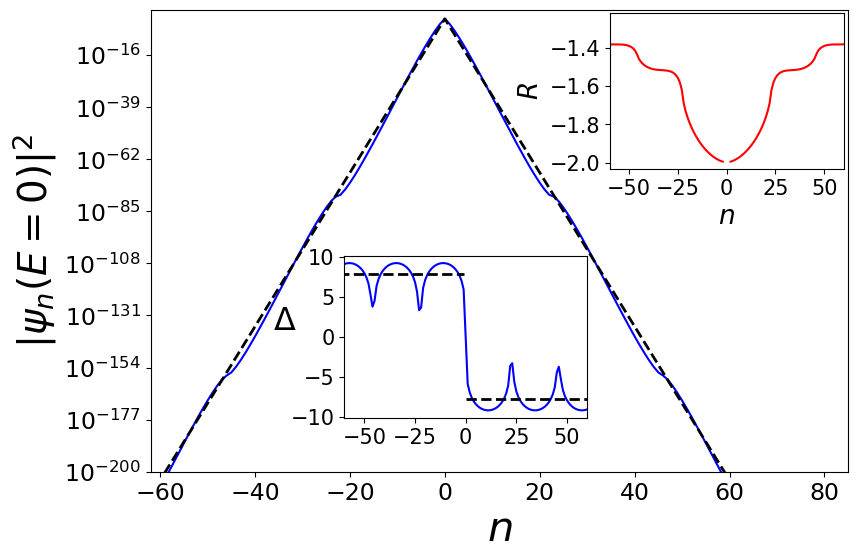}}
  \subfigure[]{\includegraphics[width = 0.9\columnwidth]{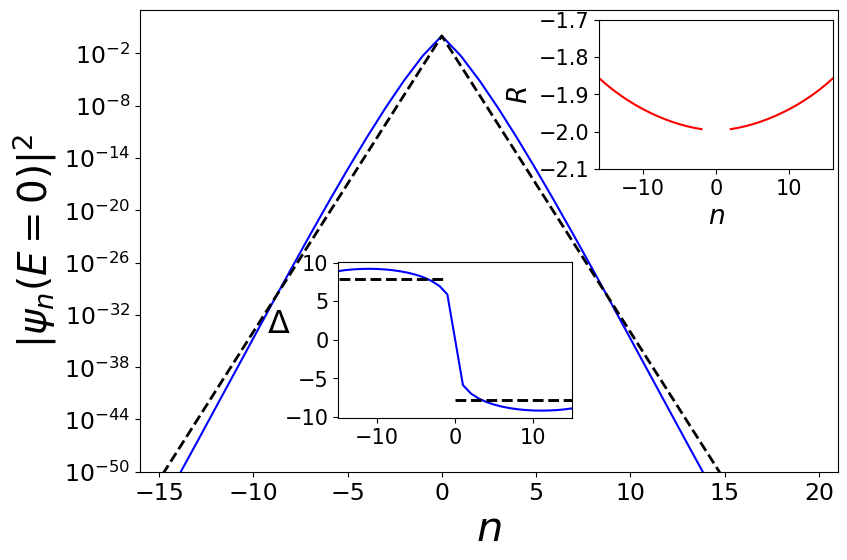}}
  \caption{
    The log-linear plot of the spatial profile (blue) of the \(E=0\) eigenstate of the Hamiltonian~\eqref{eq:full_aa_ham} with \(\wl = 47, (\alpha = 2\pi \varphi/75\)), \(\lambda = 100\), 
    \(\phi = \pi/2\), \(n_0 = 0\) and \(N=201\) sites with open boundary conditions.
    (a) The central part of the full eigenstate's profile (blue) and the exponential fit (dashed black) predicted by the theory~\eqref{eq:ll}.
    \emph{Bottom inset}: \(\lgn\), Eq.~\eqref{eq:grad_log_prob}, (blue) and the corresponding exponential fit (dashed black).
    \emph{Top inset}: \(R\) value, Eq.~\eqref{eq:ratio_log_prob}, computed for \(\mF = \lambda \alpha\).
    (b) The zoomed in central part of (a) (blue) and the exponential fit (dashed black).
    \emph{Insets}: The respective zoom-ins of the insets for \(\lgn\) and \(R\) from (a).
  }
  \label{fig:sin_ws_small_wind}
\end{figure}

We increase the winding length to \(\wl = 47\). 
The estimate of the size of the eigenstate's core~\eqref{eq:span} produces \(\msp \approx 0.3\), i.e.~the state is still strongly peaked at around a single lattice site. 
Figure~\ref{fig:sin_ws_small_wind}{\color{red}(a)} shows the log-linear plot of the probability profile for the eigenvector \(E = 0\) (blue) and the exponential fit (dashed black) predicted by the theory~\eqref{eq:ll}.
Overall the decay of the profile is exponential although significant deviations develop, that are particularly well exposed in Fig.~\ref{fig:sin_ws_small_wind}{\color{red}(b)}.
This is also confirmed by the bottom inset of Fig.~\ref{fig:sin_ws_small_wind}{\color{red}(a)}, which shows an oscillating \(\lgn\) to the left and to the right of the eigenstate's peak. 
Note that the oscillation period is \(\wl/2 \approx 23\), since the zeros of \(\sin(\alpha n)\) repeat with period \(\wl/2\).

The averaged over the oscillating regions \(\lgn\) still matches the theoretical prediction of the localization length \(\xi: 2/\xi = 2\ln\abs{\lambda/2t} = 2\ln(50) = 7.82\).
The step-function like behavior of the envelope of the cover of oscillating profile for \(\lgn\) in the inset of Fig.~\ref{fig:sin_ws_small_wind}{\color{red}(a)} implies exponential decay of the probability on average.
For short distances around the peak of the eigenstates, the profile decays super-exponentially as implied by the inset of Fig.~\ref{fig:sin_ws_small_wind}{\color{red}(b)},
where the \(R\) value in the region around \(n=0\) is computed with \(\mF=\lambda \alpha\).
The value of \(R\) is not constant and approaches \(-2\) only very close to the origin \(n=0\), 
suggesting that \(\wl=47\) is not large enough and the intermediate super-exponential localization regime only starts to emerge at around \(n=0\).

\begin{figure}
  \includegraphics[width = 0.95\columnwidth]{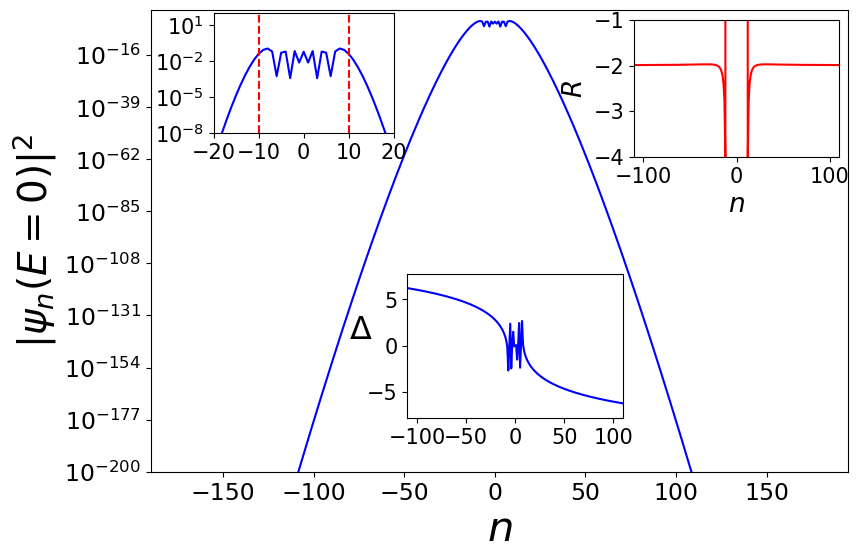}
  \caption{
    The log-linear plot of the spatial profile (blue) of the \(E = 0\) eigenstate of the Hamiltonian~\eqref{eq:full_aa_ham} with \(\wl = 3091\, (\alpha = 2\pi \varphi/5000\)), 
    all the other parameters are the same as in Fig.~\ref{fig:sin_ws_small_wind}.
    Lattice size is \(N=341\) with open boundary conditions.
    \emph{Top left inset}: zoom in around the peak of the eigenstate profile showing the \emph{plane wave} regime extending from \(-10\) to \(10\). 
    The vertical dashed red lines mark the theoretical prediction \(4|t|/|\lambda \alpha|\approx 20\).
    \emph{Top right inset}: \(R\) value~\eqref{eq:ratio_log_prob} for \(\mF = \lambda \alpha\).
    \emph{Bottom inset}: the gradient \(\lgn\)~\eqref{eq:grad_log_prob}.
  }
  \label{fig:sin_ws_large_wind}
\end{figure}

To clearly expose the intermediate lengthscale super-exponential localization, we increase the winding length up to \(\wl=3091\) keeping all the other parameters the same as before. 
Now the predicted core size becomes \(\msp=20\), and the super-exponential decay should last over one half of the winding length i.e., over \(1545\) sites (which is much larger the system size used in the numerics).
Figure~\ref{fig:sin_ws_large_wind} shows the spatial profile \(\abs{\psi_n}^2\) of the eigenstate at \(E = 0\) in the log-linear scale.
Overall the profile looks similar to the previous cases with a peak at the origin and decay away from it.
First, we note that around the peak \(n=0\) the profile does not decay at all, but rather resembles a plain wave as shown in the top left inset of Fig.~\ref{fig:sin_ws_large_wind}:
here the kinetic energy dominates over the potential energy up to the length scale \(4 \abs{t}/(\abs{\lambda \alpha})\) \(\approx 20\) for the choice of the parameters, so that the particle is almost free and the eigenstate resembles a plain wave.
This estimate is confirmed in the top left inset of Fig.~\ref{fig:sin_ws_large_wind} where the dashed red lines mark the \(4 \abs{t}/(\abs{\lambda \alpha})\) scale.
The gradient \(\lgn\) is not constant but decays (with increase of \(n\)) as shown in the bottom inset of Fig.~\ref{fig:sin_ws_large_wind}.
This suggests that the decay away from the center of the profile deviates strongly from the exponential fit~\eqref{eq:ll}.
We see from the main figure and the left inset the absence of exponential localization all the way down to \(10^{-200}\), making the expected exponential asymptotic regime absent for all practical purposes.
The \(R\) ratio (the top right inset of Fig.~\ref{fig:sin_ws_large_wind}) quickly approaches the constant value \(-2\) away from the peak, implying the onset of the super-exponential decay, i.e., the analytical prediction of Eq.~\eqref{eq:decay_prob}.

\subsection{Metallic regime: \(\abs{\lambda} < 2\abs{t}\)}

\begin{figure}
  \includegraphics[width = 0.95\columnwidth]{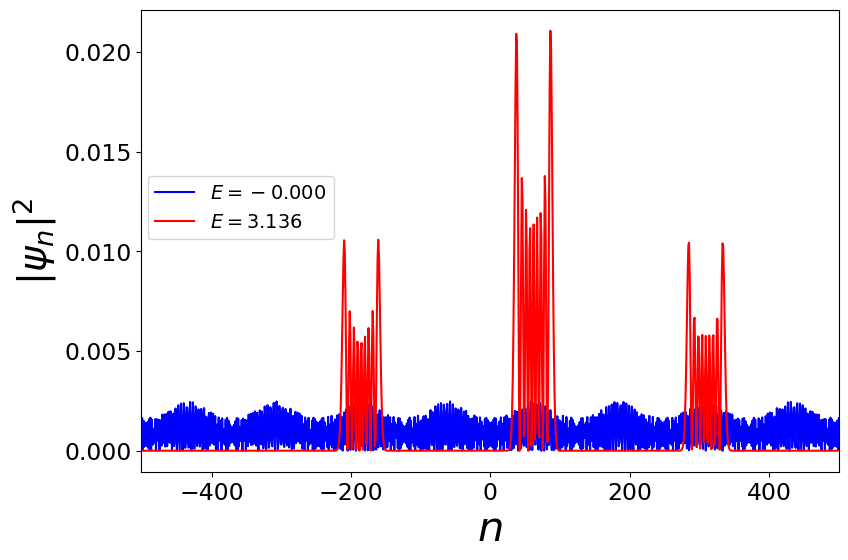}
  \caption{
    Spatial profiles of the \(E = 0\) (blue) and \(E = 3.136\) (red) eigenstates in the metallic regime \(\abs{\lambda} < 2\abs{t}\) of the Aubry-Andr\'e model \eqref{eq:full_aa_ham}. 
    The parameters of the Hamiltonian: \(\wl = 248\, (\alpha = 2\pi \varphi/400)\), \(\lambda = 1.5\), \(\phi = \pi/2\) and lattice size \(N=1001\) with open boundary conditions.
  }
  \label{fig:ext_reg_small_win}
\end{figure}

\begin{figure}
  \includegraphics[width = \columnwidth]{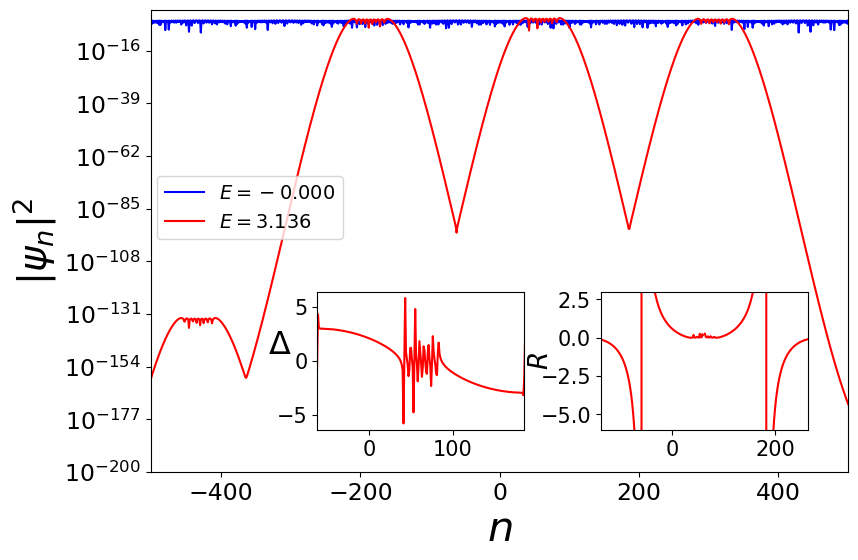}
  \caption{
    Log-linear plot of the spatial profiles of the \(E = 0\) (blue) and \(E= 3.136\) (red) eigenstates in the extended regime \(\abs{\lambda} < 2\abs{t}\) of the Aubry-Andr\'e model~\eqref{eq:full_aa_ham}.
    The parameters of the Hamiltonian: \(\wl = 248\, (\alpha = 2\pi \varphi/400\)), \(\lambda = 1.5\), \(\phi = \pi/2\) and lattice size \(N=1001\) with open boundary conditions.
    \emph{Left inset}: gradient \(\lgn\) for the \(E = 3.136\) eigenstate (red).
    \emph{Right inset}: the \(R\) value for the \(E = 3.136\) eigenstate (red).
  }
  \label{fig:ext_reg_large_win}
\end{figure}

In the metallic regime \(\abs{\lambda} < 2\abs{t}\) all eigenstates are asymptotically extended. 
For eigenenergies \(|E| < 2t\) eigenstate profiles are indeed relatively evenly spread, with only weak modulations stemming from the AA potential variations.
However, for energies closer to the band edges \(2t < |E| \leq 2t+\lambda\) the profiles of the eigenstates are different and consist of periodically arranged cores with super-exponential decay and growth in between them,
as follows from the numerical results presented below.
We choose the winding length \(\wl = 248\, (\alpha = 2\pi \varphi/400\)) and \(\lambda = 1.5\), \(\phi = \pi/2\), \(t = 1\).
The profiles of two different eigenstates located in the different parts of the spectrum---at the center, \(E=0\), and closer to the edge, \(E=3.136\),---are shown in Fig.~\ref{fig:ext_reg_small_win} and Fig.~\ref{fig:ext_reg_large_win} on a linear and logarithmic scales respectively.
At the center of the spectrum, the eigenstates are extended, as illustrated by the \(E=0\) case, which shows a modulation of the spatial profile with half the winding length \(\wl /2 = 124\).

Away from the center of the spectrum eigenstates show localization on intermediate lengthscales far into the tails of the eigenstates.
The \(E=3.136\) eigenstate illustrates this behavior. 
The non-constant, decreasing gradient \(\lgn\) in left bottom inset of Fig.~\ref{fig:ext_reg_large_win} implies a non-exponential decay/growth of the \(|\psi_n(E = 3.136)|^2\) 
(for convenience \(\lgn\) only for the central core is shown). 
The central core is located around site \(n_0 = 62\) where AA is approximated by the quadratic potential~\eqref{eq:cos_poly2} rather than the linear one~\eqref{eq:sin_poly1}.
The right bottom inset of Fig.~\ref{fig:ext_reg_large_win} shows the \(R\) value for the central core evaluated from Eq.~\eqref{eq:ratio_log_prob} with \(\mu = 2\), \(\mF = \lambda \alpha^2 / 2\).
It does never flatten around the value \(-2\) as expected for the localization induced by quadratic potential on a lattice, Eqs.~\eqref{eq:cos_poly2} and~\eqref{eq:decay_prob}.
The intermediate localization becomes stronger towards the edges of the spectrum (not shown).
Yet this localization is only a transient behaviour on intermediate lengthscales and the eigenstate eventually ``revives" for large enough system sizes, in accordance with the theory~\cite{aubry1980analyticity}.
Nevertheless for the winding lengths and the system sizes considered, some eigenstates appear as faster-than-exponentially localized for all practical purposes since their amplitudes go down all the way to \(10^{-95}\) (see Fig.~\ref{fig:ext_reg_large_win}).

\section{Intermediate super-exponential localization in discrete-time unitary maps}
\label{sec:uc}

\begin{figure}
  \includegraphics[width = 0.75\columnwidth]{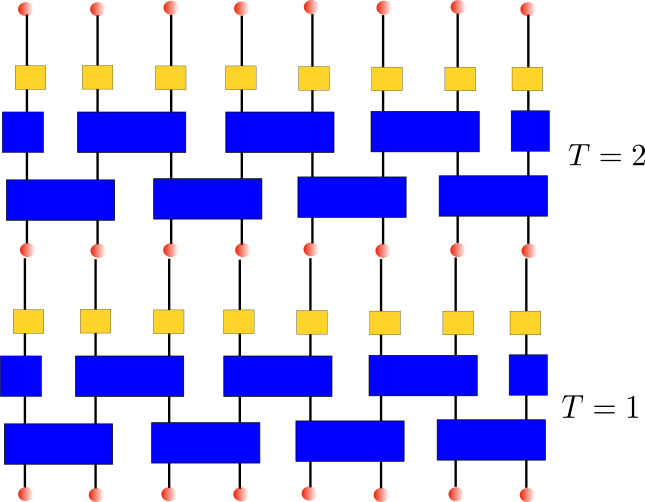}
  \caption{
    Pictorial representation of two time steps, \(T=1,2\) of the unitary map. 
    The small red spheres indicate lattice sites. 
    Blue shaded rectangles indicate the coin operations, steps I and II in Eq.~\eqref{eq:uc_split}, 
    the yellow shaded squares indicate phase operations, step III in Eq.~\eqref{eq:uc_split}, which depend on the lattice site positions.
  }
  \label{fig:uc_draw}
\end{figure}

The emergence of the super-exponential localization on intermediate length scales followed by an asymptotic exponential decay in the AA Hamiltonian~\eqref{eq:full_aa_ham} can also be realized in other settings with quasiperiodic potentials.
In particular, we consider a discrete-time unitary map, the linear and non-disordered version of the maps used in Ref.~\onlinecite{hamza2009dynamical,malishava2022lyapunov}.
These maps provide a convenient computational tool for studying long-time dynamics of nonlinear systems~\cite{vakulchyk2019wave, malishava2022lyapunov}.
It is therefore important to extend these maps to various settings, e.g.~to mimic the effect of a DC field.

We start by defining the unitary map.
For that we consider a single particle on an infinite 1D chain with time evolution governed by the brickwork action of local unitary operators and described by the following three steps depicted schematically in Fig.~\ref{fig:uc_draw}: 
All the sites are split into disjoint pairs of the n.n.~sites.\\
\underline{Step I}: 
Probability amplitudes on every pair of sites are mixed by applying an orthogonal \(\mathrm{SO}(2)\) matrix:
\begin{align*}
  & a_{2n-1}(T + 1) = \cos\theta \psi_{2n-1}(T) + \sin\theta \psi_{2n}(T), \\
  & a_{2n}(T + 1) = -\sin\theta \psi_{2n-1}(T) + \cos\theta \psi_{2n}(T);
\end{align*}
\underline{Step II}:
make a cyclic shift of all the sites by one lattice spacing and mix the amplitudes of the pairs again by applying the same \(\mathrm{SO}(2)\) matrix:
\begin{align*}
  & b_{2n}(T + 1) = \cos\theta a_{2n}(T + 1) + \sin\theta a_{2n+1}(T + 1), \\
  & b_{2n+1}(T + 1) = -\sin\theta a_{2n}(T + 1) + \cos\theta a_{2n+1}(T + 1);
\end{align*}
\underline{Step III}:
add local random phases \(\varepsilon_n\) to all the amplitudes
\begin{align}
  \label{eq:uc_split}
  \psi_n(T + 1) = e^{i \varepsilon_n} b_n(T + 1)\;.
\end{align}
Combining all the three steps together produces the final unitary map:
\begin{align}
  \psi_{2n}(T+1) & = e^{i \varepsilon_{2n}} [ \cos^2 \theta \psi_{2n}(T) - \cos\theta \sin\theta \psi_{2n-1}(T) \notag \\
  & +\sin\theta \cos\theta\psi_{2n+1}(T) + \sin^2 \theta \psi_{2n+2}(T)], \notag \\
  \psi_{2n+1}(T+1) & = e^{i \varepsilon_{2n+1}} [\cos^2 \theta \psi_{2n+1}(T) - \sin\theta \cos\theta \psi_{2n}(T) \notag \\
  & +\cos\theta \sin\theta \psi_{2n+2}(T) + \sin^2 \theta \psi_{2n-1}(T)]\;.
  \label{eq:UC_evl}
\end{align}

In the absence of the local phase, \(\varepsilon_n = 0\) for all sites \(n\), the map is translationally invariant and has extended eigenstates with eigenvalues \(e^{-iE}\) sitting on the unit circle in the complex plane.
The eigenvalues are parametrized by their angle (phase) \(E\) which is a function of lattice momentum or wave number \(k\): \(E = \pm \arccos (\cos^2 \theta + \cos k \sin^2\theta)\)~\cite{malishava2022thermalization}. 
For our work we consider \(0 < \theta < \pi\).
The bandwidth is \(\Delta E = 4\theta\). 
For a quasiperiodic local phase \(\varepsilon_n = -\alpha n\) with an irrational \(\alpha/\pi\), this unitary evolution displays asymptotic exponential localization with localization length
\begin{gather}
  \xi = \frac{1}{|\ln(|\sin \theta|)|}
  \label{eq:loclengthUM}
\end{gather}
as shown in Ref.~\cite{cedzich2021anderson}(Proposition 3.2). 
This setup is similar to the case of the electric quantum walk with an irrational phase~\cite{cedzich2013propagation, arnault2020quantum} or the aperiodic quantum walks~\cite{riberio2004aperiodic}. 

Similarly to the AA case, we consider \(\abs{\alpha} \ll 1\) so that \(\abs{\alpha n} \ll \abs{n}\) for all \(n\) on a finite chain, so that \(e^{- i \alpha n} \approx 1 - i \alpha n\) is justified.
Furthermore to allow nearest neighbor hopping only (as in the AA Hamiltonian) we also require \(\theta \ll 1\), so that \(\cos \theta \approx 1\) and \(\sin \theta \approx \theta\). 
Then the \(\alpha \theta\) and \(\theta^2\) terms are negligible compared to \(\alpha\) and \(\theta\) respectively, and the unitary map~\eqref{eq:UC_evl} becomes approximately
\begin{subequations}
  \begin{align}
    \psi_{2n}(T+1) & \approx (1 - i \alpha (2 n))\psi_{2n}(T) \\
    & - \theta (\psi_{2n-1}(T) - \psi_{2n+1}(T)), \notag \\
    \psi_{2n+1}(T+1) & \approx (1 - i \alpha (2n+1))\psi_{2n+1}(T) \\
    & - \theta (\psi_{2n}(T) - \psi_{2n+2}(T))\;. \notag
  \end{align}
  \label{sub:uc_ws}
\end{subequations}
The two equations~\eqref{sub:uc_ws} are the same up to the relabelling of the lattice sites, and correspond to the discrete time version of the 1D Wannier-Stark Schr\"odinger Hamiltonian
\begin{align}
  & \mh_\mathrm{ap}\ket{\psi(T)} = i\left[ \ket{\psi(T+1)} - \ket{\psi(T)}\right], \notag \\
  \label{eq:um_ws}
  & \mh_\mathrm{ap} = i \theta \sum_n \ketbra{n}{n+1} -  \ketbra{n}{n-1} + \alpha \sum_n n \ketbra{n}{n}\;.
\end{align}
From the above expressions we can define the effective DC field strength \(\mF = \alpha\) and hopping parameter \(t = i\theta\).
Based on this mapping of the parameters, we expect that the eigenstates of the exact unitary map~\eqref{eq:UC_evl} display super-exponential localization on intermediate length scales for \(\abs{\alpha}, \theta \ll 1\),
followed by the asymptotical exponential localization, similarly to the Aubry-Andr\'e case.
We also define the winding length \(\wl = \lceil 2\pi/\alpha\rceil\), that controls the intermediate length scale behaviour just like in the AA model above. 
In analogy to the AA chain reasoning we predict that the span of the core of an eigenstate is \(\msp \approx \frac{4\abs{\theta}}{\alpha} \approx \frac{2\abs{\theta}}{\pi}\wl\) 
and super-exponential decay should go uninterrupted over one winding length \(\wl\).

To verify these predictions numerically, we first transform the unitary map~\eqref{eq:UC_evl} into a discrete time quantum random walk~\cite{aharonov1993quantum}.
The unitary map~\eqref{eq:UC_evl} acts differently on odd and even sites, and has consequently two bands in the spectrum.
The unitary map can be viewed as a special case of the split-step discrete-time quantum walk (SS-QW) \cite{kitagawa2012topological, mallick2016dirac} on a 1D chain.
In order to reformulate our unitary map as the quantum walk, each pair of odd and even sites of the original chain are merged into a single site of a new chain with two internal states:
spin up \(\uparrow\) and spin down \(\downarrow\), corresponding to the odd and even sites of the original chain respectively~\footnote{Alternatively, one can map even and odd sites to spin up and down respectively.}
\begin{align}
  & \ket{2n-1} \mapsto \ket{m, \uparrow} = (1 ~0)^T \otimes \ket{m}, \notag \\
  & \ket{2n} \mapsto \ket{m, \downarrow} = (0~1)^T \otimes \ket{m}\;.
\end{align}
The spin degrees of freedom define the coin space on every site of the new chain.
The new lattice site index \(m=2n\) takes only even values, in order to match with the minimum distance between unit cells in the original lattice.
A single time step of the SS-QW reads:
\begin{align}
  \label{eq:ss-qw}
  & U_\mathrm{ss-qw} = P \cdot S_2 \cdot C_2  \cdot S_1 \cdot C_1, \\
  & P = \sum_{m \in 2\mz}
  \begin{pmatrix}
    e^{i\varepsilon_{m,\uparrow}} & 0 \\
    0 & e^{i\varepsilon_{m,\downarrow}} \\
  \end{pmatrix}
  \otimes \ketbra{m}{m}, \notag \\
  & S_2 = \sum_{m \in 2\mz}
  \begin{pmatrix}
    \ketbra{m+2}{m} & 0 \\
    0 & \ketbra{m}{m}\\
  \end{pmatrix}, \notag \\
  & S_1 = \sum_{m \in 2\mz}
  \begin{pmatrix}
    \ketbra{m}{m} & 0 \\
    0 & \ketbra{m-2}{m} \\
  \end{pmatrix}, \notag \\
  & C_j = \sum_{m \in 2\mz}
  \begin{pmatrix}
    \sin \theta & (-1)^j\cos \theta \\
    (-1)^{j+1}\cos \theta & \sin \theta \\
  \end{pmatrix}
  \otimes \ketbra{m}{m}\;. \notag
\end{align}
Here \(C_{1,2}\) are unitary operations in the coin space while \(S_{1,2}\) are the coin state dependent shift operators on the chain.
\(P\) is the coin and site dependent unitary phase shift.
Note that the conventional discrete-time quantum walk is recovered by setting \(C_2\) to an identity matrix.
In what follows we focus on the localization properties of eigenstates of the evolution operator of this quantum walk \eqref{eq:ss-qw}.

\subsection{Numerical results}

We consider a 1D chain with periodic boundary condition: the phase \(\varepsilon_{n,\downarrow,\uparrow}\) is not periodic and has a jump in value between the sites \(N\) and \(1\).
This choice of the boundary condition is dictated by convenience of avoiding the open boundary conditions for a discrete time quantum walk.
As we are expecting super-exponential localization, the boundary conditions should not matter much once the eigenstate is localized away from the site \(1, N\) where the jump of the phase \(\varepsilon_{n,\downarrow,\uparrow}\) occurs.
We diagonalize \(U_\mathrm{ss-qw}\), transform an eigenstate of the quantum walk~\eqref{eq:ss-qw} at a given pseudo-energy \(e^{-iE}\) to an eigenstate of the original unitary map~\eqref{eq:UC_evl}, 
and check the spacial decay of the resulting eigenstate using the same two metrics as in the AA model case: \(\lgn\) and \(R\).

\begin{figure}
  \includegraphics[width = 0.45\textwidth]{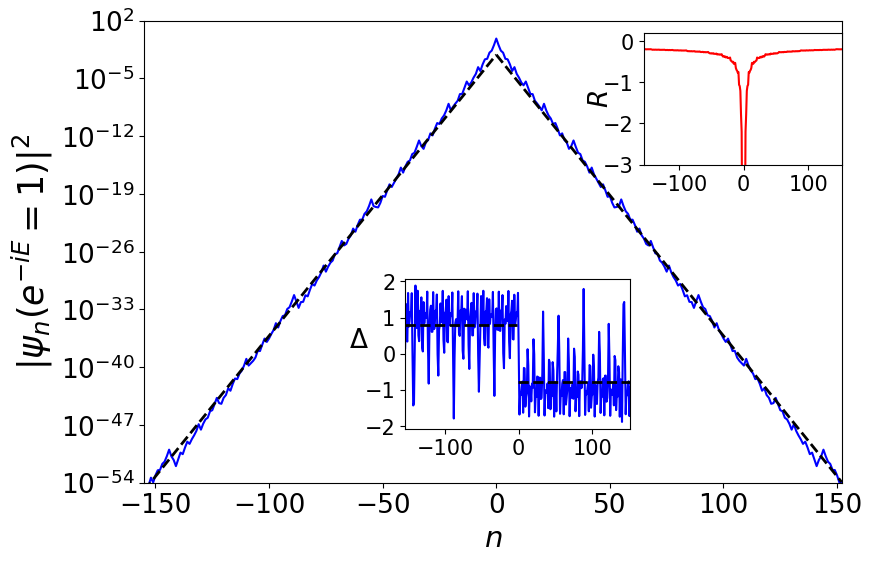}
  \caption{
    The log-linear plot of the spatial profile of the eigenstate at \(E=0\) (blue) of the unitary map with \( \alpha = \theta = 2\pi \varphi,\, \wl = 1\). 
    The lattice size is \(N = 154\) for the quantum walk.
    The dashed black lines are the exponential fit with the slope given by \(2/ (\xi) = 2/2.549 = 0.785\), computed from Eq.~\eqref{eq:loclengthUM}.
    \emph{Bottom inset}: the gradient \(\lgn\) (blue) and its average (dashed black) that is used in the exponential fit.
    \emph{Top inset}: the \(R\) value (red).
  }
  \label{fig:uc_wl_small}
\end{figure}

For small winding length, \(\wl=1\), we expect clear exponential localization.
This is confirmed in Fig.~\ref{fig:uc_wl_small}: the \(E=0\) eigenstate profile matches well the exponential fit given by~\eqref{eq:loclengthUM}, with only minor deviations.
The bottom inset shows \(\lgn\): while fluctuating, its average value agrees well with the value given by Eq.~\eqref{eq:loclengthUM}.
In the top inset the \(R\) value is different from \(-2\) indicating the absence of the super-exponential localization.

\begin{figure}
  \includegraphics[width = 0.45\textwidth]{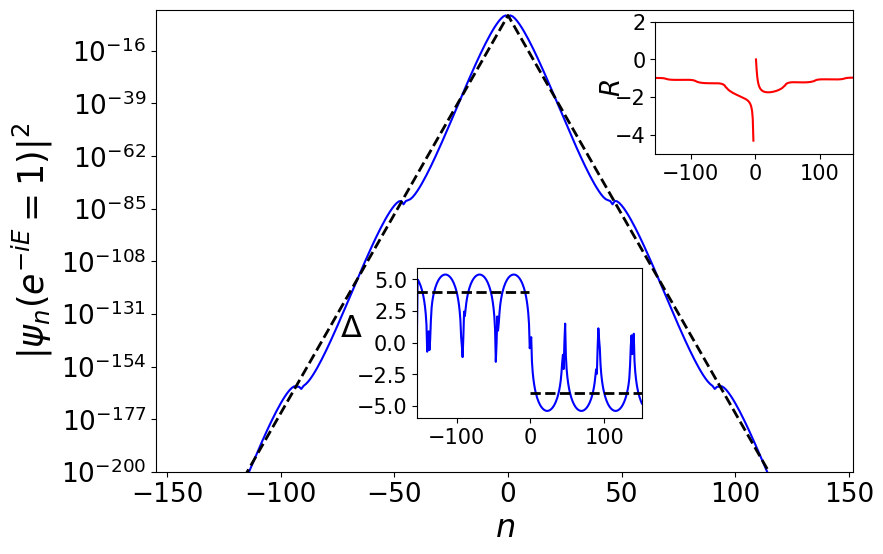}
  \caption{
    The log-linear plot of the spatial profile of the \(E=0\) eigenvector (blue) of the unitary map with \(\wl = 47,\, \alpha = 2\pi \varphi/75 = \theta\), \(\xi=0.5\) (according to the formula~\eqref{eq:loclengthUM}).
    The system size for the quantum walk is \(N = 154\).
    The dashed black lines are the exponential fits with the slope \(2/ (\xi) = 4\).
    \emph{Bottom inset}: the gradient \(\lgn\) of the profile (blue) and the gradient used in the exponential fit (dashed black).
    \emph{Top inset}: the \(R\) value (red).
  }
  \label{fig:uc_wl_int}
\end{figure}

Upon increasing the winding length, for intermediate value \(\wl = \lceil 2\pi/\alpha \rceil = 47\), the \(E = 0\) eigenstate profile (blue) is still fitted reasonably by an exponential (dashed black) as suggested by Fig.~\ref{fig:uc_wl_int}, 
however significant deviations are visible.
The bottom inset shows the gradient \(\lgn\) (blue): there are significant oscillations, however the analytical expression for the localization length~\eqref{eq:loclengthUM} (dashed black) still provides a reasonable fit to the eigenstate profile.
The top inset shows the \(R\) value: it is not equal to \(-2\) and changes with \(n\), indicating the absence of Wannier-Stark-like localization.

\begin{figure}
  \includegraphics[width = \columnwidth]{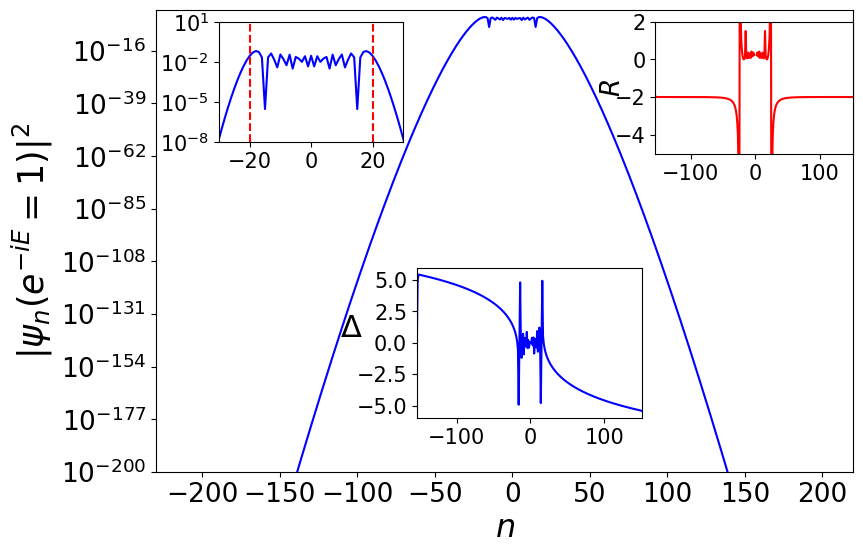}
  \caption{
    The log-linear plot of the spatial profile of the \(E=0\) eigenvector (blue) of the unitary map with \(\wl = 3091,\, \alpha = 2\pi \varphi/5000\), \(\theta = 10\alpha\) and \(\xi=0.257\).
    Lattice size for the quantum walk is \(N=154\).
    \emph{Bottom inset}: the gradient \(\lgn\) (blue) is not constant.
    \emph{Top right inset}: the \(R\) value (red) flattens around the \(-2\) value.
    \emph{Top left inset}: zoom in around the peak of the eigenstate profile showing the \emph{plane wave} regime extending from \(-20\) to \(20\). 
    The vertical dashed red lines mark the theoretical prediction \(4|\theta/\alpha| = 40\).
  }
  \label{fig:uc_wl_large}
\end{figure}

For the large winding number \(\wl = \lceil 2\pi/\alpha \rceil = 3091\), the \(E=0\) eigenstate's profile is shown in Fig.~\ref{fig:uc_wl_large}.
The profile looks very similar to that of the AA model with large winding number, shown in Fig.~\ref{fig:sin_ws_large_wind}.
Close to the peak the profile resembles a plane wave, decaying further away from the peak.
The top left inset shows the plane wave part of the profile around the peak.
The dashed red lines mark the predicted boundaries of the plane-wave-like regime, \(4|\theta/\alpha| = 40\), where kinetic and potential energy become comparable, and match well the numerics.
The bottom inset shows the gradient \(\lgn\): it is not constant over all distances shown implying the absence of the exponential localization.
At the same time the value of \(R\), shown in the top right inset, is almost constant and close to \(-2\) for large distances, 
signaling the factorial, i.e., super-exponential localization down to the extremely small values of \(\approx 10^{-200}\).
This makes the detection of the asymptotic exponential localization regime quite difficult as it has to emerge at still smaller values.

\section{Conclusion}
\label{sec:concl}

We have shown that for a general power-law potential on a tight-binding chain all the eigenstates decay factorially i.e., super-exponentially.
This is a slower decay as compared to the case in continuous space.

We further demonstrated that such super-exponential decay emerges on intermediate lengthscales in the 1D AA model.
In the insulating regime of the AA model, \(\abs{\lambda} > 2\abs{t}\), we demonstrated that by fine-tuning the parameters of the Hamiltonian one can obtain an intermediate super-exponential localization regime, 
that persists down to extremely small values of the amplitudes of the eigenstates. 
This makes the asymptotic exponential localization irrelevant for many practical purposes.
The presence of this regime is controlled by the winding length \(\wl \approx 2\pi/\alpha\) over which the AA potential almost repeats itself.
This behavior is based on the observation that for large winding lengths, much bigger than the lattice spacing or comparable to the system size, the Aubry-Andr\'e potential is well approximated by a power-law potential, 
e.g.~the Wannier-Stark (linear) or discrete harmonic oscillator (quadratic) potentials, for extended parts of the chain.

In the metallic regime, \(\abs{\lambda} < 2\abs{t}\), theory predicts extended eigenstates only.
For small \(\wl\), within our finite system size, all the eigenstates are indeed extended.
For large winding lengths \(\wl\), the eigenstates close to the center of the spectrum, \(E\lesssim 2t\), are extended.
However, surprisingly, the eigenvalues closer to the edges of the spectrum feature super-exponentially localized peaks. 
One would have to consider system sizes much larger than the winding length \(\wl\) in order to see that these state are extended.

We also demonstrated analytically and confirmed numerically that such intermediate super-exponential localization preceding the asymptotic exponential localization, is also achievable in unitary maps/discrete-time quantum walks.
This setup could be implemented in state-of-art experimental devices~\cite{alderete2020quantum}.

Our work suggests a novel regime to explore in the well-known quasi-periodic lattice model which is useful in many branches of physics. 
The large winding length parameters can be achieved by the state-of-art experimental devices, e.g.~in a one-dimensional bichromatic potential obtained by superimposing two optical lattices having very different wavelengths~\cite{modugno2009exponential}.

\begin{acknowledgments}
  The authors acknowledge the financial support from the Institute for Basic Science (IBS) in the Republic of Korea through the project IBS-R024-D1.
  The work of A.M.~was partially funded by the National Science Centre, Poland (grant  2021/03/Y/ST2/00186) within the QuantERA II Programme, 
  that has received funding from the European Union’s Horizon 2020 research and innovation programme under Grant Agreement No 101017733.
\end{acknowledgments}

\bibliography{flatband,mbl,general,ergodicity}

\end{document}